\documentstyle[aasms4,epsf]{article}

\newcommand{\sa}[0]    {${\cal L}$}

\lefthead{L. V. E. Koopmans \& C. D. Fassnacht}  
\righthead{II. A determination of H$_0$ with the {\sl CLASS} 
	   gravitational lens B1608+656}
\begin{document}

\title{A determination of H$_0$ with the CLASS 
       gravitational lens B1608+656:\\ 
       II. Mass models and the Hubble constant from lensing.}

\author{L. V. E. Koopmans} 
\affil{Kapteyn Astronomical Institute, University of Groningen, \\ 
       P.O.Box 800, NL-9700 AV Groningen, The Netherlands \\ 
       e-mail: leon@astro.rug.nl} 
\and 
\author{C. D. Fassnacht\altaffilmark{1}} 
\affil{California Institute of Technology\\ 
       105-24, Pasadena CA 91125, USA\\
       e-mail: cdf@astro.caltech.edu}
\altaffiltext{1}
{
	   Current Address:  
	   NRAO,
	   P.O.\ Box O,
	   Socorro, NM 87801.
	   Current e-mail:  cfassnac@aoc.nrao.edu
}

\accepted{:~1999 July 1}
\received{:~1999 April 7}

\begin{abstract}

We present mass models of the four-image gravitational lens system
B1608+656, based on information obtained through VLBA imaging, VLA
monitoring and {\it Hubble Space Telescope} (HST) WFPC2 and NICMOS
imaging.  A mass model for the lens galaxies has been determined that
reproduces (i) all image positions within the observational errors,
(ii) two out of three flux-density ratios within about 10\% from the
observed ratios and (iii) the model time delays within 1\% from their
observed values, given our best estimate of the Hubble parameter.

Using the time delays determined by Fassnacht et al.\ (1999a), we also
find that the best isothermal mass model gives H$_0=59^{+7}_{-6}$ km
s$^{-1}$ Mpc$^{-1}$ for $\Omega_{\rm m}=1$ and $\Omega_{\Lambda}=0.0$,
or H$_0=(65-63)^{+7}_{-6}$ km s$^{-1}$ Mpc$^{-1}$ for $\Omega_{\rm
m}=0.3$ and $\Omega_{\Lambda}=$0.0--0.7. The statistical errors
indicate the 95.4\% (2--$\sigma$) confidence interval. A systematic
error of $\pm15$ km s$^{-1}$ Mpc$^{-1}$ is estimated from a 20\%
(1--$\sigma$) uncertainty in the steepness of radial mass profile.

This cosmological determination of H$_0$ agrees well with
determinations from three other gravitational lens systems
(i.e. B0218+357, Q0957+561 and PKS1830-211), Type Ia Supernovae, the
Sunyaev-Zel'dovich effect and local determinations.  The current
agreement on H$_0$ -- within the 1--$\sigma$ statistical errors --
from four out of five gravitational lens systems (i) emphasizes the
reliability of its determination from isolated gravitational lens
systems and (ii) suggests that a close-to-isothermal mass profile can
describe disk galaxies (e.g. B0218+357 and possibly PKS1830-211),
ellipticals (e.g. B1608+656) and central cluster ellipticals
(e.g. Q0957+561).

The average of H$_0$ from B0218+357, Q0957+561, B1608+656 and
PKS1830-211, gives H$_0^{\rm GL}=69\pm7$ km s$^{-1}$ Mpc$^{-1}$ for a
flat universe with $\Omega_{\rm m}=1$ or H$_0^{\rm GL}=74\pm8$ km
s$^{-1}$ Mpc$^{-1}$ for $\Omega_{\rm m}=0.3$ and
$\Omega_{\Lambda}$=0.0--0.7. When including PG1115+080, these values
decrease to 64$\pm$11 km s$^{-1}$ Mpc$^{-1}$ and 68$\pm$13 km s$^{-1}$
Mpc$^{-1}$, respectively. The errors are the estimated 2--$\sigma$
errors on the average. The Hubble parameter from gravitational lenses
seems to agree best with local determinations of H$_0$ for a low
density universe, under the assumption that all lenses are nearly
isothermal.

\end{abstract}

\keywords{cosmology: gravitational lensing -- dark matter -- distance scale}

\section{Introduction}

Since the discovery of the first gravitational lens Q0957+561 (Walsh,
Carswell \& Weymann 1979), there has been considerable interest in
monitoring the radio and optical emission of its lens images (a quasar
at $z=1.41$), in order to find correlations between their light
curves. Strong correlations can give a time delay between the images
and definitively prove that they are lensed images of one background
object. Already in 1964 it was shown that such a time delay can be
used to constrain the Hubble parameter (H$_0$), if a good lens mass
model can be found (Refsdal 1964).

For Q0957+561 several long-term monitoring programs finally resulted
in a robust determination of the time delay (e.g. Kundi{\'c} et
al. 1997b).  Combined with the lens mass model a value of
H$_0=63\pm12$ km s$^{-1}$ Mpc$^{-1}$ (95\% confidence) was determined
(Kundi{\'c} et al. 1997b).  However, a new interpretation of the data
shed some doubt on the reliability of the confidence levels quoted for
the mass models (Barkana et al.\ 1998), and a new determination the
velocity dispersion of the cluster lens by Tonry \& Franx (1999)
increased the value of H$_0$ to 70$\pm7$ or 72$\pm$7 km s$^{-1}$
Mpc$^{-1}$ (1--$\sigma$), using the SPLS or FGS models from Grogin \&
Narayan (1996a, 1996b), respectively.  All values of H$_0$ are given
for a flat universe with $\Omega_{\rm m}$=1, if not specified
otherwise.

Three other gravitational lens systems have also yielded values of
H$_0$.  From the gravitational lens PG1115+080 (Weymann et al.\ 1980)
a value of H$_0=44\pm4$ km s$^{-1}$ Mpc$^{-1}$ (1--$\sigma$) was
determined (Impey et al.\ 1997) using the time delays found by
Schechter et al.\ (1997) and an isothermal lens mass model. The
gravitational lens system resides near a compact group of galaxies
(Kundi{\'c} et al.\ 1997a; Tonry 1998), which complicates the mass
model considerably.

The time delay in B0218+357 (e.g. Patnaik et al.\ 1993) was recently
determined by Biggs et al.\ (1999) to within 4\% accuracy (95\%
confidence).  This significantly reduced the uncertainty compared with
a previous determination by Corbett et al.\ (1996). Preliminary
modeling of this system gave H$_0=69^{+13}_{-19}$ km s$^{-1}$
Mpc$^{-1}$ (95\% confidence), using an isothermal mass model to
describe the lens galaxy (Biggs et al.\ 1999).

Very recently the time delay in PKS1830-211 (e.g. Jauncey et al.\
1991) was determined (Lovell et al.\ 1998). Combined with the detailed
mass model from Nair, Narasimha \& Rao (1993) and the source redshift
found by Lidman et al.\ (1998), a value of H$_0$=65$^{+16}_{-9}$ km
s$^{-1}$ Mpc$^{-1}$ (1--$\sigma$) was determined (Lovell et al.\ 1998;
Lovell, private communication).  When using an isothermal mass model
on the galaxy position from Nair et al.\ (1993) we estimate
H$_0$=75$^{+18}_{-10}$ km s$^{-1}$ Mpc$^{-1}$ (1--$\sigma$).

Using the same lens mass model for all five gravitational lens systems
makes a more consistent comparison of H$_0$ from these systems
possible (Schechter 1998), especially if their radial mass profile is
ill-constrained. We have chosen to use the isothermal mass model.

In sections 2 and 3 we will briefly discuss the observational status
of B1608+656 and new observations that are used in this paper.
Section 4 introduces the mass distribution and the procedure that is
used to solve for the mass model parameters and the Hubble parameter.
In section 5, the results for different mass models are presented. In
section 6, a determination of the Hubble parameter from B1608+656 is
given. In section 7 this estimate is compared with determinations from
other gravitational lens systems and in section 8 with determinations
from Type Ia Supernovae and the Sunyaev-Zel'dovich effect and the
local determination of H$_0$. The conclusions are summarized in
section 9.

The observed time delays between the lens images in B1608+656 are
presented in the companion paper by Fassnacht et al.\ (1999a;
hereafter Paper I).

\section{The four-image gravitational lens B1608+656}

The four-image gravitational lens system B1608+656 was discovered by
the {\it Cosmic Lens All Sky Survey} collaboration (CLASS; Myers et
al. 1995; Myers et al.\ 1999, in preparation) and independently by
Snellen et al.\ (1995).

The goal of CLASS is to compose a statistically complete sample of
radio-selected gravitational lenses, for which the selection effects
are well understood (Browne et al.\ 1997). For this purpose all
flat-spectrum ($\alpha$$<$0.5 with $S_{\nu}\propto\nu^{-\alpha}$) radio
sources brighter than 25 mJy at 5 GHz were selected between
declinations 0 and 75 degrees (Browne et al.\ 1997). These sources
($\approx$12,000 in total) were observed with the VLA in A-array at
8.4-GHz. Promising lens candidates were followed-up with MERLIN and/or
the VLBA. So far at least 12 new gravitational lens systems were found
(e.g. Jackson et al.\ 1998; Browne et al.\ 1997).

The lensed object in B1608+656 is the nucleus of radio galaxy at a
redshift $z_{\rm s}=1.394$ (Fassnacht et al.\ 1996) which has a radio
luminosity in the transition range between Fanaroff-Riley Class I and
II galaxies (Fanaroff \& Riley 1974, Snellen et al.\ 1995; Fassnacht et
al. 1996).  The primary lens galaxy has a redshift $z_{\rm l}=0.6304$
(Myers et al.\ 1995). Additional multi-frequency observations with
several radio telescopes (VLA, MERLIN, VLBA) and HST were done.  The
relevant results will be summarized in section 3.

Flat-spectrum radio sources are often variable. This is one of the
prerequisites for determining the time delay between lens images and
constraining H$_0$ (Refsdal 1964). The lensed object in B1608+656 is
compact and has varied $\approx$20$\%$ in flux density (8.4-GHz) on
time scales of weeks to months (Fassnacht 1997). Thus, a seven-month
monitoring campaign was started with the VLA in A- and B-arrays during
the 1996--1997 season to determine time delays between the image-pairs
(Paper I).  The obtained time delays, flux density ratios and
additional radio and optical data from B1608+656 will be used in this
paper to constrain the mass model of this gravitational lens system,
culminating in an estimate of the Hubble parameter.

\section{Data}
 
Most of the observations and data summarized in this section will be
presented in Fassnacht et al.\ (1999b, in preparation). The image light
curves, time delays and flux density ratios from the VLA 8.4-GHz
monitoring campaign are presented in the companion paper (Paper I).

\subsection{Image positions}

From deep VLBA 5-GHz observations of B1608+656 very precise relative
positions of the four lens images were obtained (Fassnacht et
al. 1999b, in preparation). The results are listed in Table 1.  
The four images are compact and show no extended 
structure on scales $\ga$1 mas.  Data reduction and image
fitting was done in the NRAO data reduction package {\sl AIPS}. The
statistically expected errors on the image positions are approximately
the FWHM of the restoring beam dived by the signal-to-noise of the
component (Table 1).

\begin{table}
\vbox{
\begin{center}
\begin{tabular}{crrrr}
\hline
\hline
Image & A & B & C & D \\
\hline
$x_i^o$ ($''$)     &  $\equiv$0.0000  & -0.7380    & -0.7446  &  +1.1284 \\
$y_i^o$ ($''$)     &  $\equiv$0.0000  & -1.9612    & -0.4537  &  -1.2565 \\
$\delta x_i$ (mas) &  0.0023 & 0.0043 &  0.0045    &  0.0107   \\
$\delta y_i$ (mas) &  0.0023 & 0.0046 &  0.0049    &  0.0124   \\
 &\\
$r_i^o=S^i_{\nu}\;/S^{\rm A}_{\nu}$ & $\equiv$1.000 & 0.490 & 0.508 & 0.172\\
$\delta r_i $ & $\equiv$0.000 & 0.020 & 0.020 & 0.020   \\
 &\\
$t_i^o$ (d) & 30 & $\equiv$0 & 36 & 76  \\
$\delta t_i$ (d) & +7/-7 & $\equiv$0 & +7/-7 & +9/-10 \\
\hline
\hline
\end{tabular}
\end{center}
\begin{small}
\caption{\small Properties of the radio lens images of B1608+656. The
positions were determined from VLBA observations (Fassnacht et
al. 1999, in preparation). The coordinate system is Cartesian with the
positive $x$-axis pointing west. The errors are the formal 1--$\sigma$
errors (FWHM of the beam divided by the signal-to-noise) from the VLBA
data. The flux density ratio and time delays were determined from the
VLA monitoring observations (Paper I). The errors on the time delays
(95\% confidence) were determined from Monte-Carlo simulations. The
errors on the flux density ratios are assumed larger than the formal
Monte-Carlo errors for reasons given in section 3.2.}  
\vspace{0.3cm}
\end{small}}
\end{table}

\subsection{Flux density ratios and time delays}

The light curves of the four lens images can give (i) the time delays,
if correlations between the light curves are found, and (ii) the
`true' flux density ratios between the lens images, after correction
for the time delays.

The exact procedure and results from the VLA 8.4-GHz monitoring
campaign of B1608+656 are described in the companion paper (Paper I).
In Table 1, we list the obtained results.  We use the flux density
ratios and time delays to constrain the lens mass models of
B1608+656. The errors on the time delays are between 12\% and 23\%
(2--$\sigma$), determined from Monte-Carlo simulations (Paper I).

The errors on the flux density ratios are about 0.002, determined from
the same Monte-Carlo simulations. Systematic differences between the
light curves, however, appear appreciably larger than the error on the
individual light-curve points, because the reduced $\chi^2$ between
the light curves is always significantly larger than unity (Paper I).
Because the flux density ratios are most sensitive to small changes in
the lens potential (e.g. milli-lensing, micro-lensing in the case of a
very compact source ($\ll$1 mas), the presence of small nearby
companions to the lens galaxies, etc.), small-scale scintillation or
systematic self-calibration errors, we will conservatively assume an
error of 0.02 on the flux density ratios, which is approximately the
scatter between the individual points of each light curve. In Koopmans
\& de Bruyn (1999, in preparation) it is shown that a few percent
variability in radio light curves at 8.4-GHz can be expected for weak
flat-spectrum radio sources, due to micro-lensing. A few-percent error
on the flux-density ratios is already a considerable improvement on 
the 20\% error used in previous models (e.g. Keeton \& Kochanek 1997).

\subsection{The lens galaxies}

\begin{table}
\vbox{
\begin{center}
\begin{tabular}{lrrrrrr}
\hline
\hline
  & $x_c$ ($''$)  & $y_c$ ($''$) & P.A. ($^\circ$) & $(b\;/a)_{\Sigma}$ \\
\hline
G1 & & & & \\
F555W     & 0.544  & -1.060 & 77$\rightarrow$84 & $\approx$0.45\\
F814W     & 0.521  & -1.062 & 81$\rightarrow$77 & 0.45$\rightarrow$0.60\\
F160W     & 0.446  & -1.063 & 85$\rightarrow$76 & -- \\
&\\
G2 & & & & \\
F555W     & -0.337 & -0.976 & --                & -- \\
F814W     & -0.293 & -0.965 & $\approx75$       & 0.12$\rightarrow$0.40\\
F160W     & -0.276 & -0.937 &     --            & -- \\
\hline
\hline
\end{tabular}
\end{center}}
\begin{small}
\caption{\small Surface brightness distribution parameters of the lens
galaxies G1 and G2.  The surface brightness centroid is given by
$(x_c, y_c)$, with an estimated error of 15 mas. The surface
brightness position angle (P.A; from north to east) and axis ratio
($(b\;/a)_{\Sigma}$) were determined in {\sl IRAF} by fitting
elliptical isophotes as function of radius to the surface brightness
distributions of G1 and G2. The range of positions angles (P.A.) and
axis ratios from the inner to the outer isophotes is indicated by the
arrow. The coordinate system is Cartesian with positive $x$-axis
pointing west.}
\end{small}
\vspace{0.3cm}
\end{table}

{\it Hubble Space Telescope} (HST) exposures were obtained of
B1608+656, both with the WFPC2 (F555W and F814W) and NICMOS (F160W). 
The F814W exposure of B1608+656 is presented in Jackson et
al. (1998) and Fassnacht et al.\ (1999b, in preparation).  The F555W and
F160W exposures will be presented in Fassnacht et al.\ (1999b, in
preparation).  

All three exposures show four optical lens images and two objects
confined in the region between the lens images. These objects remain
distinctly separated in the F555W, F814W and F160W exposures. We
assume they are two individual lens galaxies, perhaps merging (Jackson
et al.\ 1997; Fassnacht et al.\ 1999b, in preparation). Long-slit
spectroscopy with the Keck telescope along the two galaxies (G1 and
G2) shows no sign indicating different redshifts.  Another 
two--lens gravitational lens system (i.e. B1127+385) was recently
found by the CLASS collaboration (Koopmans et al.\ 1999), although it
only has two lens images, indicating that gravitational lens systems with multiple lens
galaxies are not uncommon.  All three HST exposures (F555W, F814W and
F160W) of B1608+656 can be found on the Castles web-page ({\sl
http://cfa-www.harvard.edu/glensdata/1608.html}) as well.

The relative positions (centroid) of the lens images and lens galaxies
were determined in {\sl IRAF}\footnote{IRAF (Image Reduction and
Analysis Facility) is distributed by the National Astronomy
Observatories, which are operated by the Association of Universities
for Research in Astronomy under cooperative agreement with the
National Science Foundation.} with respect to the brightest optical
image (A). The optical image positions agree with the radio positions
to within the observational errors of $\approx$5 mas. The lens galaxy
positions and their errors are listed in Table 2. The galaxy centroids
change as function of wavelength (i.e. between F555W, F814W and
F160W), possibly suggesting that G1 and G2 are interacting dynamically
(see section 5.3). The galaxy positions are used as additional
constraints on the lens mass model.

The surface brightness distributions of G1 and G2 were fitted with
elliptical isophotes in {\sl IRAF}. The range of axis ratios and
position angles of the elliptical isophotes are listed in Table 2.

\section{Modeling}

\subsection{The lens mass model}

We describe the surface mass distribution of the two lens galaxies
with the parameterized elliptical isothermal mass model from Kormann,
Schneider \& Bartelmann (1994). The surface density distribution of
these models is given by
\begin{equation}
        \Sigma(\xi_1,\xi_2)=\frac{\sigma^2}{2G} \frac{\sqrt{f}}{\sqrt{\xi^2 +
\xi_{\rm c}^2}},
\end{equation}
where $\xi^2=\xi_{1}^2+f^2 \xi_{2}^2$, $\sigma$ is a measure of the
line-of-sight velocity dispersion, $\xi_{\rm c}$ is the lens core
radius and $f=(a\;/b)_{\Sigma}$ is the surface density axis ratio. 
All lengths and positions will from now on be given in arcseconds.
The deflection angles and shear components of this mass model are 
also given in Kormann et al.\ (1994). It is 
simple to transform the mass distribution and corresponding 
deflection and shear fields to any required position and position 
angle.  

We will refer to the (non-)singular isothermal ellipsoidal case as the
(NIE) SIE mass distribution. We assume $\Omega_{\rm m}=1$ and
$\Omega_{\Lambda}=0$ in a smooth FRW universe, if not specified
otherwise. In section 5.4 we discuss non-isothermal mass models.

\subsection{$\chi^2$ minimization}

We use a $\chi^2$-minimization procedure to search for the free parameters
(mass model parameters, source position and Hubble parameter) that
best reproduce the observed image positions, flux density ratios and
time delays (Table 1). The $\chi^2$-function that is minimized is given by
\begin{eqnarray}
        \chi^2& = & \sum_{i=1}^4 \left[\frac{(x_{i}^{o}-x_{i}^{m})^2}
                {\delta x_{i}^2} + \frac{(y_{i}^{o}-y_{i}^{m})^2}
                {\delta y_{i}^2} \right] + \nonumber\\
              &   & \sum_{i=1}^4 \left[\frac{(r_{i}^{o}-r_{i}^{m})^2}
                {\delta r_{i}^2} \right] + \nonumber   \\
              &   & \sum_{i=1}^4 \left[\frac{(t_{i}^{o}-t(h)_{i}^{m})^2}
                {\delta t_{i}^2} \right].         
\end{eqnarray}
The first term on the right-hand side of the equation gives the
`goodness-of-fit' between the observed, $(x,y)_{i}^{o}$, and model,
$(x,y)_{i}^{m}$, image positions, where $\delta (x,y)_{i}$ are the
1--$\sigma$ errors on the observed image positions.  The second term
gives the goodness-of-fit between the observed, $r_{i}^{o}$, and
model, $r_{i}^{m}$, flux density ratios. Because the flux density of
the lensed object is a free parameter, we normalize the flux density
of the brightest observed and model image ($i=1$) to unity.  The third
term gives the goodness-of-fit between the observed, $t_{i}^{o}$, and
model, $t_{i}^{m}$, time delays. The error on the observed time delay
is given by $\delta t_{i}$. We normalize the time delay of the leading
image to zero. The model time delay is a function of the free
parameter $h=$H$_0$/(100 km s$^{-1}$ Mpc$^{-1}$). Thus, minimizing
$\chi^2$ gives not only the mass model parameters and source position,
but also a measure of the Hubble parameter, H$_0$.

A continuous Simulated Annealing Downhill-Simplex algorithm is used
to minimize $\chi^2$ (Press et al.\ 1992).  Although relatively slow,
this method is robust and allows simple adjustment of $\chi^2$ during
the minimization. It also has a high probability of finding the
global $\chi^2$--minimum compared with faster methods that need the
gradients of a complex multi-dimensional $\chi^2$--function. The latter
is of great importance if we search for the minimum--$\chi^2$ solution
in a space of many free parameters ($\ga10$).

\section{Results}

\begin{table}
\hspace{-2cm}
\vbox{
\begin{center}
\begin{tabular}{lcccccccccc}
\hline
\hline 
 Model & $\sigma$ (km s$^{-1}$) & $(b\;/a)_{\Sigma}$ & $\theta_{\rm PA}$ ($^\circ$) &
  $r_{\rm c}$ ($''$) & $\Delta t^{({\rm A-B})}$ & $\Delta t^{({\rm C-B})}$ 
    & $\Delta t^{({\rm
    D-B})}$ & H$_0^{\chi}$  & $\chi^2$ & $\chi^2$/NDF\\
\hline
F555W-I & 253.1, 201.1 & 0.78, 0.45 & 63.4, 51.2 & 0.000, 0.000 & $18.9\; h^{-1}$ &
     $22.5\; h^{-1}$ & $47.8\; h^{-1}$ & 62.5 & 48.1 & 9.6\\
F555W-II & 253.1, 201.1 & 0.78, 0.45 & 63.4, 51.2 & 0.000, 0.000 & $18.9\; h^{-1}$ &
     $22.5\; h^{-1}$ & $47.8\; h^{-1}$ & 62.5 & 15.6 & 3.9\\
F555W-III & 248.8, 210.7 & 0.80, 0.50 & 61.5, 52.0 & 0.000, 0.036 & $18.3\; h^{-1}$ & 
     $21.8\; h^{-1}$ & $45.8\; h^{-1}$ & 60.1 & 44.2 & 11.1\\
F555W-IV & 246.9, 214.9 & 0.81, 0.51 & 60.6, 52.3 & 0.000, 0.052 & $18.0\; h^{-1}$ &
     $21.4\; h^{-1}$ & $44.9\; h^{-1}$ & 59.0 & 9.4 & 3.1\\
&\\
F814W-I & 248.2, 207.2 & 0.88, 0.39 & 71.9, 53.1 & 0.000, 0.000 & $19.2\; h^{-1}$ &
     $23.2\; h^{-1}$ & $45.7\; h^{-1}$ & 61.6 & 159.7 & 31.9\\
F814W-II & 248.2, 207.2 & 0.88, 0.39 & 71.9, 53.1 & 0.000, 0.000 & $19.2\; h^{-1}$  &
     $23.2\; h^{-1}$ & $45.7\; h^{-1}$ & 61.6 & 53.5 & 13.4\\
F814W-III & 240.1, 223.8 & 0.93, 0.47 & 63.0, 54.8 & 0.000, 0.058 & $18.2\; h^{-1}$ &
     $21.9\; h^{-1}$ & $42.4\; h^{-1}$ & 57.7 & 141.2 & 35.3\\
F814W-IV & 233.3, 237.2 & 0.96, 0.52 & 36.3, 55.9 & 0.000, 0.105 & $17.3\; h^{-1}$ &
     $20.7\; h^{-1}$ & $39.4\; h^{-1}$ & 54.2 & 22.4  & 7.5\\
&\\
F160W-I   & 249.5, 207.1 & 0.90, 0.31 & 122.0, 53.8 & 0.000, 0.000 &
     $19.6\; h^{-1}$ & $24.2\; h^{-1}$ & $41.0\; h^{-1}$ & 59.7 & 622.7 & 124.5\\
F160W-II  & 249.5, 207.1 & 0.90, 0.31 & 122.0, 53.8 & 0.000, 0.000 &
     $19.6\; h^{-1}$ & $24.2\; h^{-1}$ & $41.0\; h^{-1}$ & 59.7 & 172.0 & 43.0\\
F160W-III  & 236.8, 230.8 & 0.87, 0.42 & -25.3, 57.3 & 0.000, 0.073 &
     $18.9\; h^{-1}$ & $22.9\; h^{-1}$ & $37.7\; h^{-1}$ & 57.7 & 478.9 & 119.7\\
F160W-IV  & 216.8, 266.6 & 0.70, 0.54 & -13.8, 61.1 & 0.000, 0.183 &
     $17.2\; h^{-1}$ & $20.5\; h^{-1}$ & $29.2\; h^{-1}$ & 48.1 & 19.2 & 6.4\\
\hline
\hline
\end{tabular}
\end{center}
\begin{small}
\caption{\small Isothermal mass model parameters of G1 and G2 and
determinations of H$_0$. Listed are the velocity dispersion
($\sigma$), surface density axis ratio ($(b\;/a)_{\Sigma}$), position
angle ($\theta_{\rm PA}$; north to east) and core radius ($r_{\rm
c}$). The first value is for G1, the second for G2. The time delays
$\Delta t$ (in days) are the values determined from the mass model.
H$_0^\chi$ (in km s$^{-1}$ Mpc$^{-1}$) was determined through the
minimization of $\chi^2$ as discussed in section 4.2. The last two
columns show the minimum--$\chi^2$ and reduced--$\chi^2$ values,
respectively.  We assume $\Omega_{\rm m}=1$ and $\Omega_{\Lambda}=0$
in this table.}
\vspace{0.3cm}
\end{small}}
\end{table}

We investigate several different isothermal and non-isothermal mass
models in an attempt to reproduce the observed properties of B1608+656
(Table 1).  The two lens galaxies (G1 and G2) are fixed at their
observed positions (Table 2). The HST exposures were done using
different filters (F555W, F814W and F160W), each giving slightly
different galaxy positions. Models with the lens galaxies fixed at the
positions determined from each filter are therefore considered.

It should also be noted that the resulting model flux density ratio of
image D ($r_{\rm D}$) is the only parameter never close to the
observed value, throughout all mass models. The magnification depends
critically on a combination of second order derivatives of the lens
potential and small perturbations can therefore significantly change
flux ratios. The delays and positions directly depend on zero-th and
first order derivatives of the lens potential, respectively. Hence,
they are less sensitive to perturbations by small masses (e.g. Mao \&
Schneider, 1998). Both the delay and position of image D are in
excellent agreement with observations.

For each filter we construct four mass models: (I) SIE+SIE, using all
constraints, (II) SIE+SIE, without $r_{\rm D}$ as a constraint, (III)
SIE+NIE, using all constraints and (IV) SIE+NIE, without $r_{\rm D}$
as a constraint. Mass models V and VI, presented in section 5.4, are
non-isothermal.

\subsection{SIE+SIE mass model}

The first attempt is to model both G1 and G2 with a SIE mass distribution.
In total there are 14 constraints (8 from the image positions, 3 from
the flux density ratios and 3 from the time delays) and 9 free 
parameters (source position, Hubble parameter, velocity dispersions, 
surface density axis ratios and position angles of G1 and G2). 
The number of degrees of freedom (NDF) is five. This should give
a well constrained set of free parameters, if they are non-degenerate.

Using all constraints, we minimize $\chi^2$ for the three sets of
galaxy positions (Table 2). The resulting model parameters,
minimum-$\chi^2$ and Hubble parameters are listed in Table 3 (model
I).  Using different starting values for the model parameters and a
slowly decreasing `temperature' in the Simulated Annealing algorithm
(see Press et al.\ (1992) for a more detailed description of this
minimization procedure), we ensure a very high probability of ending
up at the global $\chi^2$--minimum.

The minimum--$\chi^2$ value is significantly smaller when the HST
WFPC2 F555W galaxy positions are used, in comparison with those from
the WFPC2-F814W or NICMOS-F160W observations.
 
The $\chi^2$--minimization is repeated, this time without the use of
the flux density ratio of image D as a constraint.  The resulting
model and Hubble parameters (Table 3; models II) are similar to models
I, but $\chi^2$ has decreased more than expected on the basis of the
decreased number of degrees of freedom (NDF=4). The flux density ratio
of image D contributes disproportionally to $\chi^2$ compared with the
other constraints, possibly due to a local perturbation of the lens
potential (e.g. milli- or micro-lensing).

Lens model F555W-II has the smallest $\chi^2$ value of all SIE+SIE
mass models in Table 3. Figure 1 (left) shows both the critical-curve
and caustic structure of this model and its time-delay surface. When
moving a source from outside the caustic structure -- e.g. where the
source is singly imaged -- the number of lens images changes by two,
each time a caustic is crossed. We therefore expect the SIE+SIE models
to show seven images, of which two are highly de-magnified, because
they are located near the surface density singularities of G1 and G2.

However, the fifth image should be visible. It is formed on the saddle
point of the time-delay surface near galaxy G2 (Fig. 1).  Its
magnification is around 0.2. This should make it visible
($\ga$50--$\sigma$ detection) in the deep VLA A-array observations
available to us (Paper I). The image, however, is not detected, which
poses a strong restriction on the allowed set of models.

\begin{figure}
\plottwo{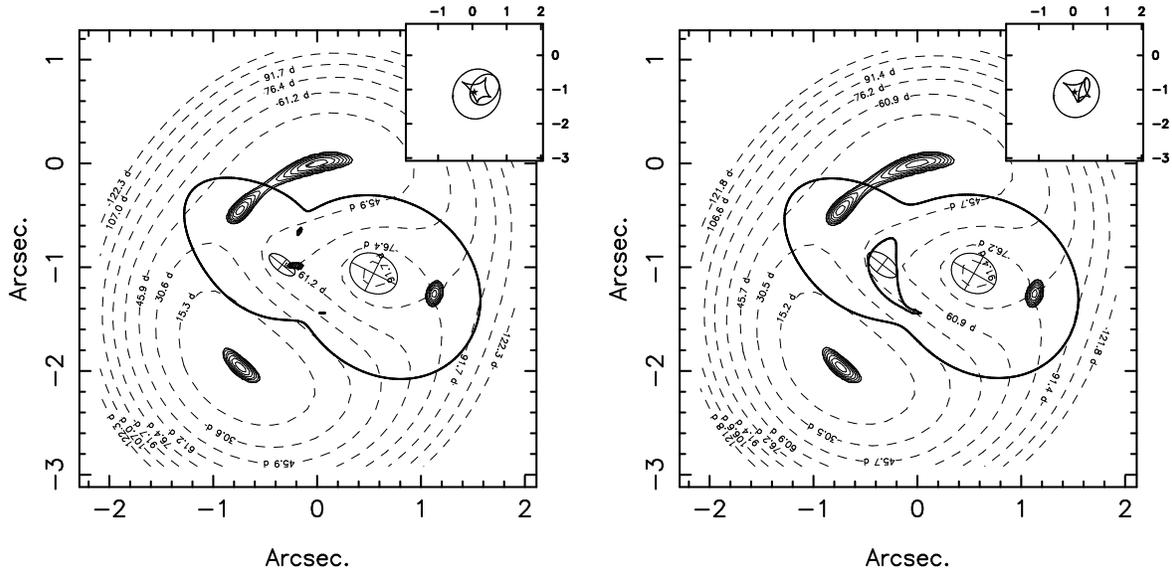}{Koopmans.fig1b.epsi}
\figcaption{\small Left: Time-delay surface (dashed lines), critical curves
(thick lines) and caustic structure (sub-panel) of the SIE+SIE lens
model F555W-II. The lens images are the projections onto the image
plane of a Gaussian shaped source (star) with a FWHM of 0.1
arcsec. The cross-haired ellipses indicate the position, axis-ratio
and position angle of G1 and G2. Right: Idem, but for the SIE+NIE mass
model F555W-IV.}
\end{figure}

\subsection{SIE+NIE mass model}

The primary reason to introduce a small core radius for G2 is to
remove the fifth image formed at the saddle-point of the time-delay
surface between G1 and G2 (Fig. 1). If the core radius is large
enough, there are no extrema on the time-delay surface and thus no
images will form between G1 and G2 (e.g. Schneider et al.\ 1992).

We minimize $\chi^2$ again for the three sets of galaxy positions.
The core radius of G2 is left free (NDF=4; models III). We repeat this
without $r_{\rm D}$ as a constraint (NDF=3; models IV). The resulting
model parameters and Hubble parameters are listed in Table 3.  Lens
model F555W-IV is shown in Fig.1 (right). The recovered image
positions, flux density ratios and time delays from model F555W-IV are
listed in Table 5.

A significant improvement in the reduced $\chi^2$ is obtained with
models IV over models II (no core radius for G2) and models I and III
(with $r_{\rm D}$ as constraint). There is also a decrease in the
reduced $\chi^2$ from models F160W-IV to F555W-IV.  This suggests that
the mass centers of G1 and G2 are more closely associated with the
emission centroids from the HST F555W exposure.

The source now only crosses two caustics, when moving it from outside
the caustic structure to its model position shown in Fig.1
(right). Thus five images instead of seven are formed. This solves the
`central-image' problem, even though no lower limit on the core radius
of G2 was imposed.  One of the five images forms near the surface
density singularity of G1 and is highly de-magnified, because the
magnification is almost inversely proportional to the surface density
close to the singularity (Kormann, Schneider \& Bartelmann 1994). Only
four observable lens images remain, as is required by the
observational constraints.

\subsection{Models with constrained position angles}

Comparing the mass model position angles of G1 and G2 (Table 3), with
the observed surface brightness position angles (Table 2), we see a
systematic $\approx$20 degree difference.

We attempted several models with position angles for G1 and G2 in the
range indicated in Table 2. The lens galaxy positions were not
constrained and they were allowed to have a finite core radius.  Even
though the system is close to being under-constrained, no satisfactory
solutions with position angles close to those listed in Table 2 could
be obtained (i.e. $\chi^2\gg10^{6}$).

The $\approx$20 degree difference in position angle between surface
brightness and surface density of G1 and G2 therefore remains. There
are several possible solutions: (i) The lens galaxies (G1 and G2) have
tri-axial halos, producing a misalignment between surface brightness
and the projected surface density of the dark matter halo (Keeton,
Kochanek, \& Seljak 1997). (ii) A dynamical interaction between G1 and
G2 could introduce a physical misalignment between the halo and the
luminous matter.  A `merger' scenario could explain why G1 and G2 look
like post-starburst galaxies (Myers et al.\ 1995) and have much bluer
colors than expected for ellipticals (Keeton, Kochanek, \& Falco 1998;
Fassnacht et al.\ 1999b, in preparation). Star-formation associated
with the deepest part of the lens potential would also explain why the
galaxy positions from the HST F555W exposure seem most closely
associated with the lens mass centers, although extinction by dust
could make this argument more complicated.  (iii) External mass
components can create the presence of an external shear near G1 and
G2, which introduces an apparent misalignment of surface brightness
and density position angles (Keeton, et al.\ 1997). B1608+656 is part
of a small group of galaxies (Fassnacht et al.\ 1999b, in
preparation). However, based on their luminosities, these galaxies do
not seem massive enough to cause significant perturbations of the lens
potential. Moreover, we have added an external shear to our best
isothermal mass model (F555W-IV; Table 3) and find no decrease in
$\chi^2$, even though NDF decreases by two.  (iv) The group of
galaxies (including G1 and G2) is associated with inter-galactic
gas. This is often seen in Hickson compact groups (Hickson 1982)
through X-ray emission (e.g. Pildis, Bregman \& Evrard 1995). Although
the group of galaxies associated with B1608+656 is not very compact or
massive, some gas could still be associated with it. If this gas
`halo' is misaligned with G1 and G2, the inferred surface density
position angles can deviate from the surface brightness position
angles, if the gas is not accounted for in the mass model.

In the cases (iii) and (iv), it remains peculiar that all image
positions, time delays and two out of the three flux density ratios
are reconstructed relatively well by the best mass model
(F555W-IV). This is hard to understand if a large-scale mass component
or external shear is missing from the mass model. We therefore prefer
either a tri-axial halo for G1 and G2 or a misalignment of the halo
and the luminous matter due to dynamical interactions of G1 and
G2. Both can explain the misalignment in position angle without
invoking new mass components.

\subsection{Non-isothermal mass models}

To investigate the validity of using an isothermal mass distribution,
we also tested models with surface density profiles given by
\begin{equation}
        \Sigma(\xi_1,\xi_2)=
        \frac{\Sigma_0}{\left[1+\left(\frac{\xi^2} {\xi_{\rm
        c}^2}\right)\right]^{-\gamma}}.
\end{equation} 
The normalization of $\Sigma_0$ is chosen to agree with that of
eqn. (1) for $\gamma=1/2$ (the isothermal model). The mass inside some
arbitrary radius diverges for profiles with $\gamma>1/2$ and $\xi_{\rm
c}=0$. We therefore place a lower limit of $10^{-3}$ arcsec on the
core radii of both G1 and G2. Both galaxies are placed on the observed
F555W galaxy positions, which seem to best represent their mass
centers (section 5.1).  We use the code developed by Barkana (1998) to
calculate the deflection angle and magnification for models with
$\gamma\ne 1/2$.

\begin{table}
\vbox{
\begin{center}
\begin{tabular}{lccccccccc}
\hline
\hline
 Model   & $\gamma$ & $\sigma$ (km s$^{-1}$) & $(b\;/a)_{\Sigma}$ & $\theta_{\rm PA}$ ($^\circ$) &
  $r_{\rm c}$ ($''$) & H$_0^{\chi}$  & $\chi^2$ \\
\hline
F555W-V &  0.3 & 293.2, 220.7 & 0.91, 0.67 & -10.8, 56.5 & 0.152, 0.017 & 27.4 &
  5.8\\
  & 0.4 & 274.5, 210.9 & 0.96, 0.57 & 36.1, 53.2 & 0.095, 0.001 & 41.8 &
  1.4\\
  & 0.5 & 269.6, 214.4 & 0.91, 0.53 & 54.7, 52.0 & 0.148, 0.078 & 50.5 &
  4.6\\
  & 0.6 & 271.0, 222.4 & 0.90, 0.51 & 55.2, 51.6 & 0.231, 0.150 & 54.7 &
  7.2\\
  & 0.8 & 294.1, 230.1 & 0.92, 0.48 & 49.3, 50.6 & 0.462, 0.249 & 56.0 &
  12.3\\
  & 1.0 & 304.1, 243.2 & 0.93, 0.48 & 41.9, 50.5 & 0.599, 0.344 & 57.4 &
  19.8\\
&\\
F555W-VI &  0.3 & 270.0, 229.4 & 0.92, 0.68 & -5.4, 56.7 & 0.003, 0.001 & 29.9 &
  195.6 \\
  & 0.4 & 261.4, 216.1 & 0.94, 0.58 & 45.4, 53.6 & 0.007, 0.005 & 44.1 &
  98.0 \\
  & 0.5 & 250.6, 207.1 & 0.80, 0.48 & 62.3, 51.7 & 0.001, 0.022 & 60.9 &
  45.2 \\
  & 0.6 & 237.0, 209.8 & 0.65, 0.44 & 64.3, 51.4 & 0.018, 0.075 & 74.8 &
  37.0\\
  & 0.8 & 237.8, 211.2 & 0.55, 0.36 & 66.0, 50.5 & 0.118, 0.150 & 89.9 &
  66.5\\
  & 1.0 & 239.5, 216.3 & 0.50, 0.33 & 66.7, 50.2 & 0.191, 0.211 & 98.5 &
  105.8\\
\hline
\hline
\end{tabular}
\end{center}
\begin{small}
\caption{\small Mass model parameters of G1 and G2 for different mass
profiles and determinations of H$_0$. Listed are the profile parameter
$\gamma$, velocity dispersion ($\sigma$), surface density axis ratio
($(b\;/a)_{\Sigma}$), position angle ($\theta_{\rm PA}$; north to
east) and core radius ($r_{\rm c}$). The first value is for G1, the
second for G2. The Hubble parameter H$_0$ (in km s$^{-1}$ Mpc$^{-1}$)
was determined through the minimization of $\chi^2$ (section 4.2). The
last column shows the minimum-$\chi^2$ value. We assume $\Omega_{\rm
m}=1$ and $\Omega_{\Lambda}=0$ in this table.}
\vspace{0.3cm}
\end{small}}
\end{table}

\subsubsection{Constraints on the radial mass profile}

Using all observational constraints, we minimize $\chi^2$ as described
in sections 5.1--3 (model F555W-VI).  The mass distributions of G1 and
G2 are given the same value of $\gamma$. We repeat this without flux
ratio $r_{\rm D}$ as a constraint (model F555W-V). The model
parameters, Hubble parameters and minimum-$\chi^2$ values are listed
in Table 4.

The table shows that $\chi^2$ minimizes in the range $\gamma$
$\approx$ 0.4--0.6, taking both models F555W-V and F555W-VI into
account. The use of $r_{\rm D}$ as constraint (model F555W-VI)
increases the $\chi^2$ values by a factor of about 10. If we assume
that $r_{\rm D}$ is correct and one of the other three images has an
incorrect flux density ratio, we still find a comparable increase in
$\chi^2$. So for all models in this range of mass profiles, $r_{\rm
D}$ is the only observable that can not be reasonably matched to its
model value.  The inclusion of $r_{\rm D}$ as a constraint changes
$\gamma$ of the best model from about 0.4 to 0.6. The lowest values of
$r_{\rm D}$ that we find is 0.24 for model F555W-VI with
$\gamma=0.6$. Thus, keeping in mind that $r_{\rm D}$ could be larger
than 0.17 -- preferred by all models (independent of $\gamma$) -- the
value of $\gamma=0.6$ can be overestimated.

When $\gamma$<$0.5$, the position angle of G1 rapidly decreases until
a prolate halo is preferred with $\theta_{\rm PA, G1}\approx-(5-11)$
degrees for $\gamma=0.3$. Values of $\gamma$ much smaller than 0.5 are
therefore on the basis of comparison with the surface brightness
positions angle not very likely.  Moreover, for $\gamma\le0.4$ all
minimum--$\chi^2$ models form a fifth central image, similar to the
SIE+SIE models (section 5.1). This is not the case for $\gamma\ge0.5$.

The range of $\gamma$$\approx$0.4--0.6 for G1 and G2 agrees with
constraints on $\gamma$ from the gravitational lenses Q0957+561
(Grogin \& Narayan 1996a; Barkana et al.\ 1998) and MG1654+1346
(Kochanek 1995). The models for both of these prefer values close to
$\gamma$=0.5 (i.e. isothermal).

\subsubsection{The radial mass profile and H$_0$}

In Figure 2, we plot H$_0$ versus $\gamma$. For models F555W-V, H$_0$
increases from 27 to 57 km s$^{-1}$ Mpc$^{-1}$ between $\gamma=0.3$
and 1.0. For models F555W-VI, H$_0$ increases from about 30 to 99 km
s$^{-1}$ Mpc$^{-1}$.

\begin{figure}
\plotone{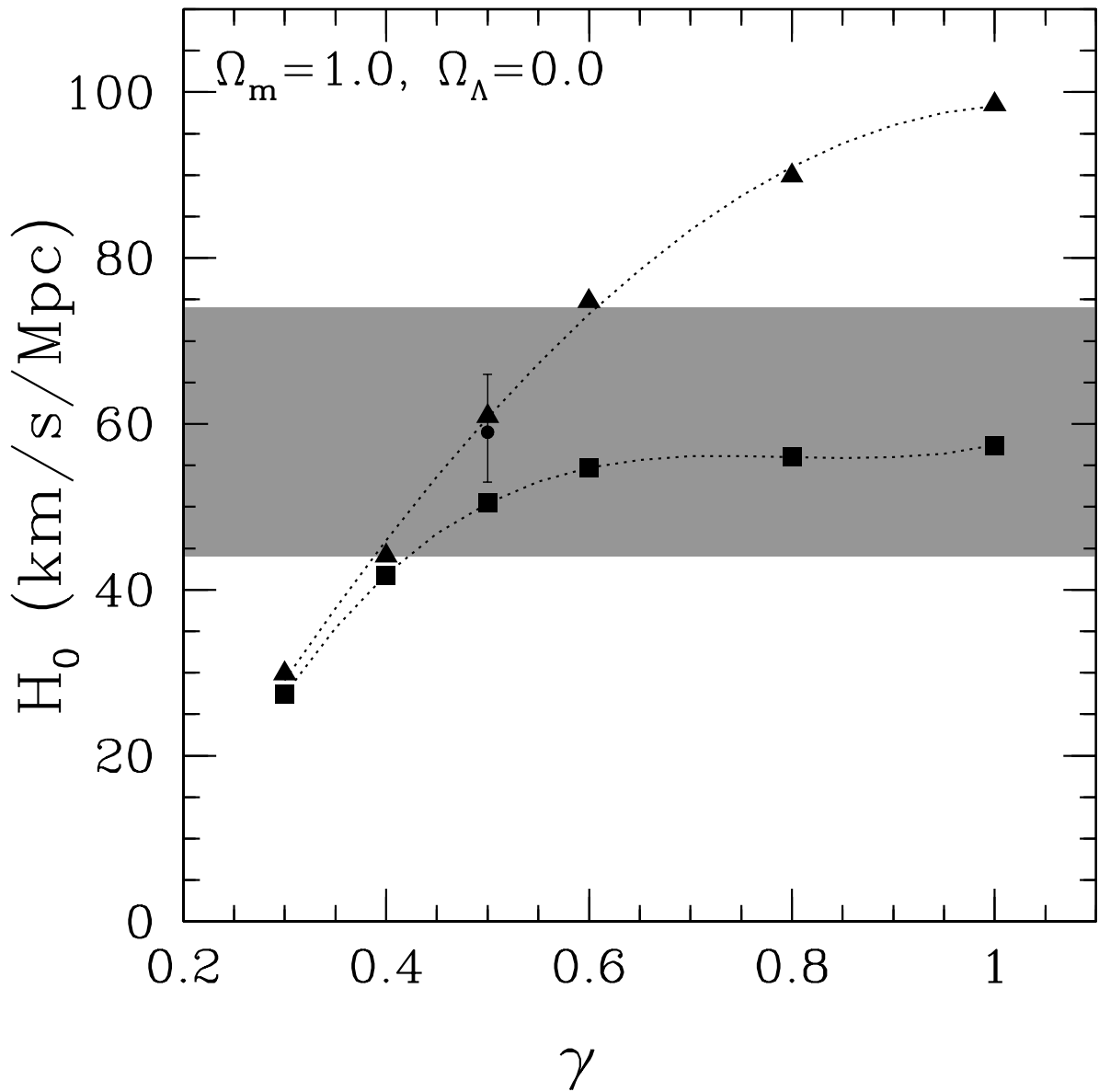}
\figcaption{\small The Hubble parameter H$_0$ from B1608+656 plotted against the
profile parameter $\gamma$ for the minimum-$\chi^2$ mass model.  The
triangles indicate the optimum value of H$_0$ obtained from the
$\chi^2$ minimization with all flux density ratios as constraints.
The squares indicate the optimum value of H$_0$ when omitting the flux
density ratio of image D. The shaded region indicates the range of
H$_0$ when $\gamma=0.50\pm0.10$ (section 5.4.1). This introduces a
systematic error of $\pm15$ km s$^{-1}$ Mpc$^{-1}$.  The filled dot
and the 2--$\sigma$ statistical error bars at $\gamma=0.5$ indicate
the value of H$_0$ from the `best' isothermal mass model (F555W-IV;
Table 3; section 6). The dotted lines are a third-order polynomial
fit.}
\end{figure}

Because models F555W-V are not constrained by $r_{\rm D}$, the core
radius of G1 can rapidly increase with $\gamma$ (Table 4).  This gives
an overall flatter mass profile of G1 inside the Einstein radius
compared with larger radii ($\gg \xi_{\rm c}$), reducing the growth of
H$_0$ for $\gamma>0.5$. However, larger core radii of G1 also give
values for $r_{\rm D}\ga0.5$, which is outside any reasonable range.
In models F555W-VI the growth of $\xi_{\rm c}$ is suppressed by the
small value of $r_{\rm D}$.

Not only do B1608+656 and Q0957+561 have similar mass profiles, they
also give similar values of H$_0$ as we will see later on. In the rest
of the paper we assume that both G1 and G2 are isothermal. The value
of $\gamma$ is conservatively taken as $0.50\pm0.10$ for the reasons
given in section 5.4.1. The models for $\gamma=0.3,0.5$ and 1.0 (Table
4) are shown in Figure 3.

\begin{figure}
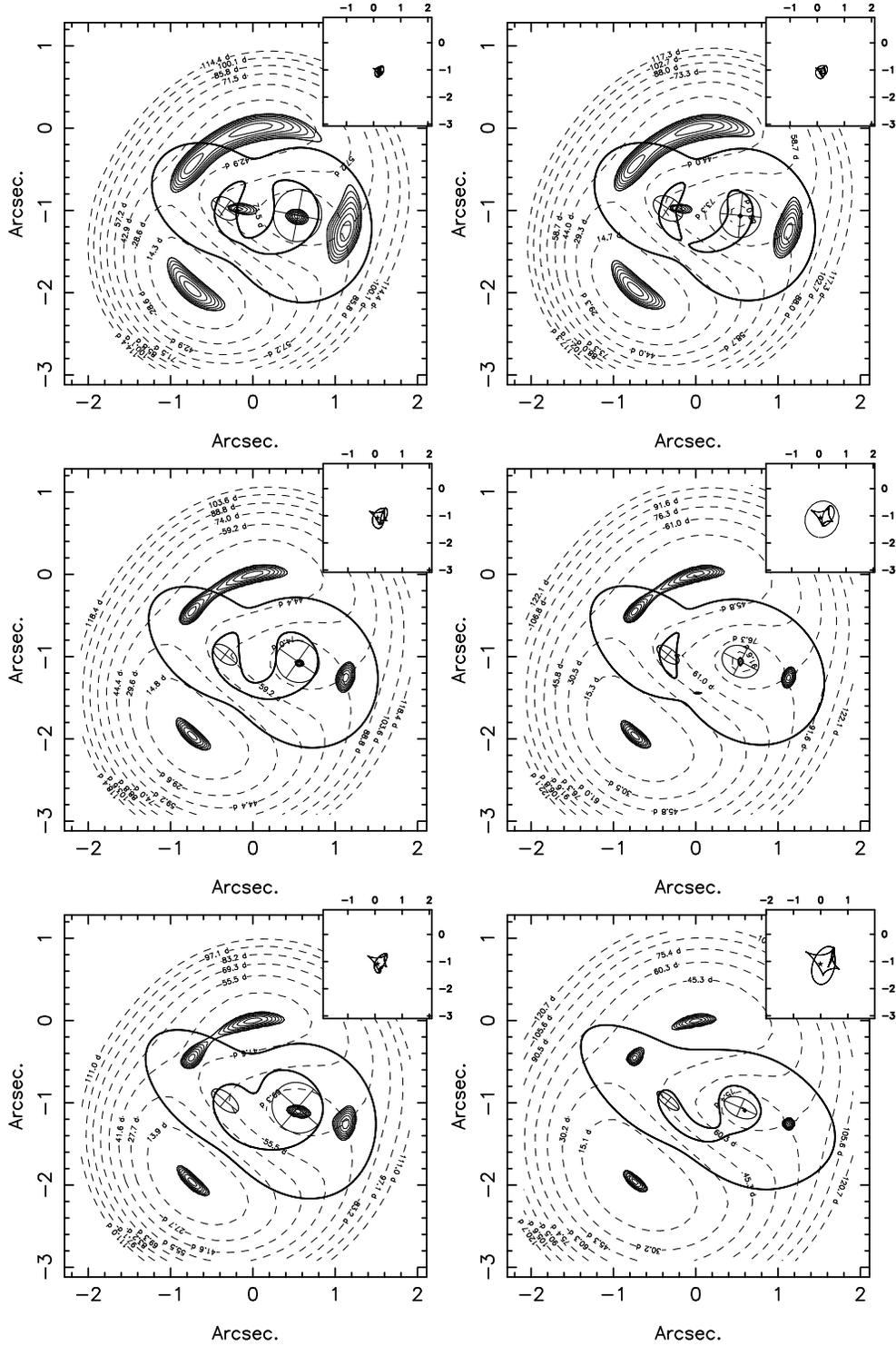

\begin{center}
\leavevmode
\vbox{%
  \hbox{%
  \epsfysize=6.5cm \epsffile{Koopmans.fig3a.epsi}
  \epsfysize=6.5cm \epsffile{Koopmans.fig3b.epsi}}
  \hbox{%
  \epsfysize=6.5cm \epsffile{Koopmans.fig3c.epsi}
  \epsfysize=6.5cm \epsffile{Koopmans.fig3d.epsi}}
  \hbox{%
  \epsfysize=6.5cm \epsffile{Koopmans.fig3e.epsi}
  \epsfysize=6.5cm \epsffile{Koopmans.fig3f.epsi}}}
\figcaption{\small Same as Fig.~1, now for models F555W-V and
F555W-VI. The figures on the left (upper to lower) show models F555W-V
for $\gamma=0.3,0.5$ and 1.0, respectively. The figures on the right
are for models F555W-VI.}
\end{center}
\end{figure}

\section{The Hubble parameter from B1608+656}

The best estimate of the Hubble parameter (i.e. H$_0^{\chi}$) from
B1608+656 ranges from 60 to 63 km s$^{-1}$ Mpc$^{-1}$ for the
SIE+SIE mass models. This range increases to 48 -- 60 km s$^{-1}$
Mpc$^{-1}$ for the SIE+NIE mass models (Table 3). All values of H$_0$
are given for a flat universe with $\Omega_{\rm m}$=1, if not
specified otherwise.

If we only consider models with (i) relatively low reduced $\chi^2$
values ($\la 10$) and (ii) models that do not give a fifth observable
image, then only models IV (Table 3) remain viable candidates. The
best isothermal model, F555W-IV, gives H$_0^{\chi}=59$ km s$^{-1}$
Mpc$^{-1}$ and a very smaller scatter ($\approx$1\%) between the values of
H$_0$ from the individual time delays. This small scatter is a
coincidence and not a result from the $\chi^2$--minimization. We have
tried different values for the delays, which gave proportionally
different values of H$_0$.  Even without the these additional three
constraints, model F555W-IV remains the best model (i.e. smallest
$\chi^2$).

We will therefore take H$_0=59$ km s$^{-1}$ Mpc$^{-1}$ to be our
best estimate of the Hubble parameter from B1608+656, given that G1
and G2 can be modeled with an isothermal mass distribution. Model
F555W-IV reproduces all observables within an acceptable range, except
for the flux density ratio of image D. It does not produce a bright
central image as required by its absence in the observations. It is
the only plausible isothermal model that cannot be excluded at very
high confidence level, based on its $\chi^2$ (assuming the flux
density ratio of image D is indeed perturbed). The surface density
position angles of G1 and G2 are also within an acceptable range from
the surface brightness position angles, considering that G1 and G2
could be merging (Jackson et al.\ 1997; Fassnacht et al.\ 1999b, in
preparation).  All other models in Table 3 can be rejected on the
basis of one of these criteria.

Clearly, an uncertainty remains regarding the explicit assumption of
the isothermal mass distribution for G1 and G2. This is reflected in
the systematic error on H$_0$, given in section 6.1.  However, the use
of similar isothermal mass models makes a direct comparison with other
lens systems possible, if their radial mass profile is ill-constrained
(Schechter 1998; section 7).

\begin{table}
\vbox{
\begin{center}
\begin{tabular}{lccccc}
\hline
\hline
Image & A & B & C & D \\
\hline
$x_i^m$ ($''$) & 0.0000 & -0.7380 & -0.7446 &  +1.1284 \\
$y_i^m$ ($''$)  & 0.0000 & -1.9612 & -0.4537 & -1.2565 \\
&\\
$r_i^m$ &  $\equiv$1.00 & 0.55 & 0.51 & 0.29\\
$\mu_i$ & 6.09 & 3.36 & -3.11 & -1.78 \\
&\\
$t_i^m$ (d) & $18.0\; h^{-1}$ & 0.0 & $21.4\; h^{-1}$ & $44.9\; h^{-1}$  \\
\hline
\hline
\end{tabular}
\end{center}
\begin{small}
\caption{\small Image properties as determined from our best model
(F555W-IV; Table 3). Except for $r_{\rm D}$ all observables are well
recovered (see Table 1). The best estimate of $h$ is 0.590 from model
F555W-IV. The image magnification and parity is given by $\mu_i$. The
source position is (0.0459$''$,-1.0774$''$).}
\end{small}}
\vspace{0.3cm}
\end{table}

\subsection{The errors on H$_0$ from B1608+656}
 
To estimate the statistical error on H$_0$ found from model F555W-IV, we
vary it over a range of fixed values and minimize $\chi^2$. 
We determine the range within which the minimum-$\chi^2$ value 
increases by less than 4.0. This range indicates the 95.4\% or 2--$\sigma$
(statistical) confidence range of H$_0$ (Press et al.\ 1992). We find:
$$ 
  {\rm H}_0^{\small 1608}=59^{+7}_{-6}\mbox{ km s}^{-1}\mbox{ Mpc}^{-1}
     \;\;(\Omega_{\rm m}=1,\;\Omega_{\Lambda}=0).
$$ 
We regard this as the `best' estimate of H$_0$ determined from the
available observational constraints of B1608+656, under the explicit
assumptions mentioned previously.

The $\approx$0.1 error on $\gamma$ introduces a systematic error in
H$_0$ of $\pm15$ km s$^{-1}$ Mpc$^{-1}$ (see Fig.2). This error
incorporates the much smaller uncertainties on H$_0$ as a result of
the chosen galaxy positions, core radii or sets of constraints (Tables
3 and 4).

\section{H$_0$ from other gravitational lens systems}

Four other gravitational lens systems have yielded values for H$_0$ at
present.  We will discuss them separately and compare the results with
those from B1608+656. All values of H$_0$ are given for a flat
universe with $\Omega_{\rm m}$=1, if not specified otherwise.

(i) The first discovered gravitational lens system Q0957+561 (Walsh et
al. 1979) recently yielded H$_0=63\pm12$ km s$^{-1}$ Mpc$^{-1}$ (95\%
confidence) from a time delay of 417$\pm$3 days (95\% confidence)
found from optical light curves (Kundi{\'c} et al.\ 1997b). Although
less constrained, Haarsma et al.\ (1999) found a delay consistent
with this from radio light curves. The error on H$_0$ is dominated by
uncertainties in the model of the mass distribution of the lens
(Kundi{\'c} et al.\ 1997b). Barkana et al.\ (1998) showed that the error
on H$_0$ could be somewhat larger than found by Kundi{\'c} et
al. (1997b), due to increased uncertainties in the assumed mass
model. Moreover, a new determination of the velocity dispersion of the
cluster lens by Tonry \& Franx (1999) indicates H$_0$=70$\pm7$ or
72$\pm7$ km s$^{-1}$ Mpc$^{-1}$ (1--$\sigma$), using the SPLS or FGS
models from Grogin \& Narayan (1996a, 1996b), respectively. The
surface density profile of the dominant lens mass distribution in
Q0957+561 was shown to be close to isothermal (Grogin \& Narayan
1996a; Barkana et al.\ 1998).

(ii) The second gravitational lens system discovered, PG1115+080
(Weymann et al.\ 1980), recently gave a value of H$_0=44\pm4$ km
s$^{-1}$ Mpc$^{-1}$ (1--$\sigma$) using an isothermal mass model
(Impey et al.\ 1997) and the time delays found by Schechter et
al.\ (1997).  This value is low compared with most determinations of
H$_0$, not only those from gravitational lensing. The lens mass
consists of an elliptical galaxy and a nearby galaxy group.  For
truncated halo models or constant M/L-ratio models, H$_0$ could go up
to as much as $65\pm5$ km s$^{-1}$ Mpc$^{-1}$. However, the large
B-band M/L-ratios ($\ge$13) suggests that dark matter is present.
This makes constant M/L models unlikely, if the luminous and dark
matter distributions are different (Impey et al.\ 1997).  For H$_0$ to
be $\ga60$ km s$^{-1}$ Mpc$^{-1}$, the dark matter halo has to be
truncated just outside the Einstein radius. This suggests that the
lens is almost completely stripped of its halo, possibly by
interaction with the nearby compact group (Kundi{\'c} et al.\ 1997a;
Tonry 1998; Impey et al.\ 1997).  PG1115+080 remains a complicated
system with large uncertainties, mostly due to the uncertain
contribution of the surface density of the group at the position of
the primary lens galaxy.  The lack of strong radio emission from the
lens images limits the information that can be obtained on the mass
distribution through deep radio or VLBI observations
(e.g. Q0957+561). However, the lensed quasar forms an optical Einstein
ring around the lens, providing valuable additional information about
the lens potential.

(iii) Recently, a robust time delay of 10.5$\pm$0.4 days (95\%
confidence) was determined between the two compact radio images in the
radio Einstein ring B0218+357 (Biggs et al.\ 1999). Preliminary
modeling based on the VLBI structure in both images (Patnaik et
al. 1995) yielded H$_0=69^{+13}_{-19}$ km s$^{-1}$ Mpc$^{-1}$ (95\%
confidence), using an isothermal mass model (Biggs et
al. 1999). B0218+357 is an `isolated' spiral galaxy lens and has no
apparent nearby massive companions.

(iv) The gravitational lens PKS1830-211 consists of an extended source
lensed into an Einstein ring (e.g. Jauncey et al.\ 1991).  The bright
compact core of the flat-spectrum source is lensed in two images
embedded in the ring structure. The source is highly variable
(e.g. Lovell et al.\ 1998) and can therefore be used to determine the
time delay between the two images. Van Ommen et al.\ (1995) found a
time delay of 44$\pm$9 days, derived from very poorly sampled light
curves. However, recently an unambiguous delay of 26$^{+4}_{-5}$ days
was found by Lovell et al.\ (1998).  With the mass model from Nair et
al. (1993) and the recently determined source redshift $z_{\rm
s}$=2.507$\pm$0.002 (Lidman et al.\  1998; private communication), this
delay yields H$_0$=65$^{+16}_{-9}$ km s$^{-1}$ Mpc$^{-1}$
(1--$\sigma$) (Lovell et al.\ 1998).  To be able to compare the
determination of H$_0$ from PKS1830-211 with those from the other four
gravitational lens systems, we model the lens galaxy as a singular
isothermal mass distribution (Kormann, Schneider \& Bartelmann
1994). We place the mass distribution at the galaxy position
determined by Nair et al.\ (1993) and minimize the difference between
the observed and model image positions and flux density ratio.  We
then find H$_0$=75$^{+18}_{-10}$ km s$^{-1}$ Mpc$^{-1}$ (1--$\sigma$).  We
will use this value in the rest of the paper.

\subsection{The average of H$_0$ from B0218+357, Q0957+561,
  PG1115+080, B1608+656 and PKS1830-211}

There is excellent agreement between H$_0$ from B0218+357, Q0957+561,
B1608+656 and PKS1830-211 (from now on called sample \sa; see
Fig. 4). The rms scatter in H$_0$ from these four systems is
$\approx$10\%, comparable to the 1--$\sigma$ statistical error on
H$_0$ from the individual gravitational lens systems. This strongly
suggests that (i) systematic effects between these four systems are
relatively small and (ii) some systematic effect remains in
PG1115+080.  Is this small number statistics or are the quoted
systematic errors on H$_0$ overestimated?  If the systematic errors
were not correlated between the different lens systems, a scatter of
some 20 per cent could be expected.

The common factor between {\it all} five gravitational lens
systems, however, is the use of the elliptical isothermal mass
model. The systematic error on H$_0$ is dominated by deviations from
this mass profile. The rms scatter in H$_0$ therefore suggests that a
`universal' mass profile can reasonably well describe the spiral
galaxies in B0218+357 and PKS1830-211, the two (elliptical) galaxies
in B1608+656 and the cluster elliptical in Q0957+561. Moreover, both
B1608+656 and Q0957+561 suggest that this profile is close to
isothermal (section 5.4.1).

Extra constraints on the radial mass profile will therefore have the
largest impact on the reliability of H$_0$ from these lenses.  In
B0218+357 and PKS1830-211, constraints on the radial mass profile can
be obtained through a detailed analysis of the radio Einstein ring
(e.g. Kochanek \& Narayan 1992; Biggs et al.\ 1999), in B1608+656
through analysis of the optical arcs and in Q0957+561 through
observations of additional sources in the field and new radio
structure (e.g. Barkana et al.\ 1998).  Also in PG1115+080,
observations of the optical Einstein ring could provide valuable new
information on the lens potential (e.g. Impey et al.\ 1998).

Thus, under the explicit assumption that the mass models are
isothermal, the errors on H$_0$ appear dominated by statistical
errors. We therefore take the average of H$_0$ (with equal weights)
from sample \sa. We find
$$ 
     {\rm H}_0^{\rm GL}=69\pm7 \mbox{ km s}^{-1}\mbox{ Mpc}^{-1}
     \;\;(\Omega_{\rm m}=1,\; \Omega_{\Lambda}=0),
$$ 
where the error is the 2--$\sigma$ error on the average (i.e
not the scatter). We have used the latest determination of H$_0$ from
Q0957+561 by Tonry \& Franx (1999).  If we add PG1115+080 to the
\sa~sample, this value decreases to 64$\pm$11 km s$^{-1}$ Mpc$^{-1}$
(2--$\sigma$).

The \sa~sample of gravitational lens systems gives H$_0$ within 10\%
accuracy at a 2--$\sigma$ confidence level. It also excludes the value
from PG1115+080 with $\ga$4--$\sigma$ (see section 8 and Table
6). However, the systematic errors on this value of H$_0$ remain around
20 per cent and it is crucial to reduce this through additional
observations, which can pin down their precise radial mass profiles.

\section{A comparison between Gravitational lens, SNIa, S-Z and local 
         determinations of H$_0$}

\begin{table}
\vbox{
\begin{center}
\begin{tabular}{cccc}
\hline
\hline
   Method  & $\Omega_{\rm
     m}$ & $\Omega_\Lambda$ & H$_0$\\
\hline
   1  & 1.0 & 0.0  & 64$\pm$11 \\
   1  & 0.3 & 0.0--0.7 & 68$\pm$13 \\ 
   2  & 1.0 & 0.0 & 69$\pm$7 \\
   2  & 0.3 & 0.0--0.7 & 74$\pm$8 \\
   3  & 1.0 & 0.0 & 64$\pm$7 \\
   3  & 0.3 & 0.0--0.7 & 67$\pm$9 \\
   4  & 1.0 & 0.0 & 67$\pm$6 \\
   4  & 0.3 & 0.0--0.7 & 70$\pm$7 \\
\hline
\end{tabular}
\end{center}}
\begin{small}
\caption{\small The average of H$_0$ from local and cosmological
determinations. The error indicates the 2--$\sigma$ error on the 
averages. The methods are: (1) All lenses (i.e. B0218+357, 
Q0957+561, PG1115+080, B1608+656 and PKS1830-211), (2) all lenses,
except for PG1115+080, (3) all methods (i.e. all lenses, SNe Ia, S-Z 
and local) and (4) all methods, except for PG1115+080.}
\end{small}
\vspace{0.3cm}
\end{table}

To determine H$_0$ on cosmological scales one can also make use of
Type Ia Supernovae or the Sunyaev-Zel'dovich (SZ) effect.

In principle the S-Z effect is very powerful method to determine
H$_0$, but systematic effects, such as cluster elongation and
clumpiness, are poorly understood. Most measurements give relatively
low values of H$_0$. However, X-ray selection minimizes the effects of
cluster elongation. From a sample of X-ray selected clusters Myers et
al. (1997) recently found H$_0$=54$\pm$14 km s$^{-1}$ Mpc$^{-1}$
(1--$\sigma$).
 
The best determination of H$_0$ from Type-Ia Supernovae comes from the
High-Z Supernovae Search Team (Riess et al.\  1998). Based on 50 SNe-Ia
events, they find ${\rm H}_0^{\rm SNIa}=65\pm7\mbox{ km s}^{-1}\mbox{
Mpc}^{-1}$ (1--$\sigma$; independent of $\Omega_{\rm m}$ and
$\Omega_{\Lambda}$). The error includes uncertainties on the
calibration of the SNIa absolute magnitudes and the zeropoint of the
Cepheid distance scale (e.g. Riess et al.\ 1998). Their results also
seem to (i) rule out $\Omega_{\rm m}=1$ for a flat universe at
$\ga7$-$\sigma$ confidence and (ii) indicate $\Omega_{\Lambda}>0$ at
$\ga3$-$\sigma$ confidence (e.g. Riess et al.\ 1998). However, the
determination of H$_0$ from SNe Ia is a `standard-candle' method.  It
can be influenced by a host of systematic effects, such as evolution,
extinction, selection biases and weak lensing (Riess et al.\ 1998).
Moreover, the determination depends on the Cepheid distance scale and
is therefore not independent of the local determinations.  This value
should therefore be regarded with some caution.

So far, we have only regarded H$_0$ from gravitational lenses in a
flat universe with $\Omega_{\rm m}$=1. But evidence has been mounting
over the last few years supporting a low density universe with
$\Omega_{\rm m}$=0.2--0.3 (e.g. Carlberg et al.\ 1996; Bahcall, Fan \&
Cen 1997; Riess et al.\ 1998).

In a low-density universe with $\Omega_{\rm m}$=0.3 and
$\Omega_{\Lambda}$ ranging from 0.0 to 0.7, the Hubble parameter
determined from gravitational lenses increases by about 7\%, depending
on the precise lens and source redshifts. We then find
$$  {\rm H}_0^{\rm 1608}=(65-63)^{+7}_{-6} \mbox{ km
    s}^{-1}\mbox{Mpc}^{-1}
$$ for $\Omega_{\rm m}$=0.3 and $\Omega_{\Lambda}$=0.0--0.7, with
2--$\sigma$ errors. The average from sample \sa~ becomes
$$
  {\rm H}_0^{\rm GL}=74\pm8\mbox{ km s}^{-1}\mbox{ Mpc}^{-1}
  \;\;(\Omega_{\rm m}=0.3,\; \Omega_{\Lambda}=0.0-0.7), 
$$ 
with 2--$\sigma$ errors (including the range in
$\Omega_{\Lambda}$). If we add PG1115+080 to the \sa~sample, this
value decreases to 68$\pm$13 km s$^{-1}$ Mpc$^{-1}$. For a low
density universe with $\Omega_{\rm m}=0.3$, a robust value of
H$_0^{\rm GL}$ is found which only weakly depends on
$\Omega_{\Lambda}$ ($\la2\%$). It is also in good agreement with the
determination from Type Ia Supernovae.

Moreover, the cosmological gravitational lens determination of H$_0$
agrees best with the local determination of H$_0=73\pm6$ (statistical)
$\pm11$ (systematic) km s$^{-1}$ Mpc$^{-1}$ ({\it Hubble Space
Telescope} Key Project for the Extra-galactic distance Scale; Freedman
et al.\ 1998) in a low density universe. Once both the local and
gravitational lens determinations of H$_0$ are well constrained
(i.e. small statistical and systematic errors), the agreement between
both values can be used to constrain the density of the universe and
possibly the cosmological constant (e.g. Refsdal 1966; Kayser \&
Refsdal 1983). However, if the local determinations of H$_0$ are
systematically larger than the determinations on cosmological scales,
this could suggest that we live in an under-dense part of the universe
(e.g. Zehavi et al.\ 1998).

We have plotted the local determination of H$_0$ from Freedman et al.
(1998) and those determined on cosmological scales in Figure 4. It
shows that all determinations of H$_0$ agree with those from sample
\sa~to within 1--$\sigma$, except for PG1115+080.  To be complete, we
have listed the averages of H$_0$ from different combinations of
methods and cosmologies in Table 6. It shows a maximum of 10 km
s$^{-1}$ Mpc$^{-1}$ difference between several determinations of the
average of H$_0$. We have included the determination of H$_0$ from
PG1115+080 in methods 1 and 3 (see Table 6), although it lies
$\ga$4--$\sigma$ outside any range listed in Table 6. The probability of
having one 4--$\sigma$ outliers in a sample of five is $\ll 1$\%,
making it unlikely that this value is due to a statistical fluke only.
 
\begin{figure}
\begin{center}
\leavevmode
\vbox{%
\epsfysize=12cm \epsffile{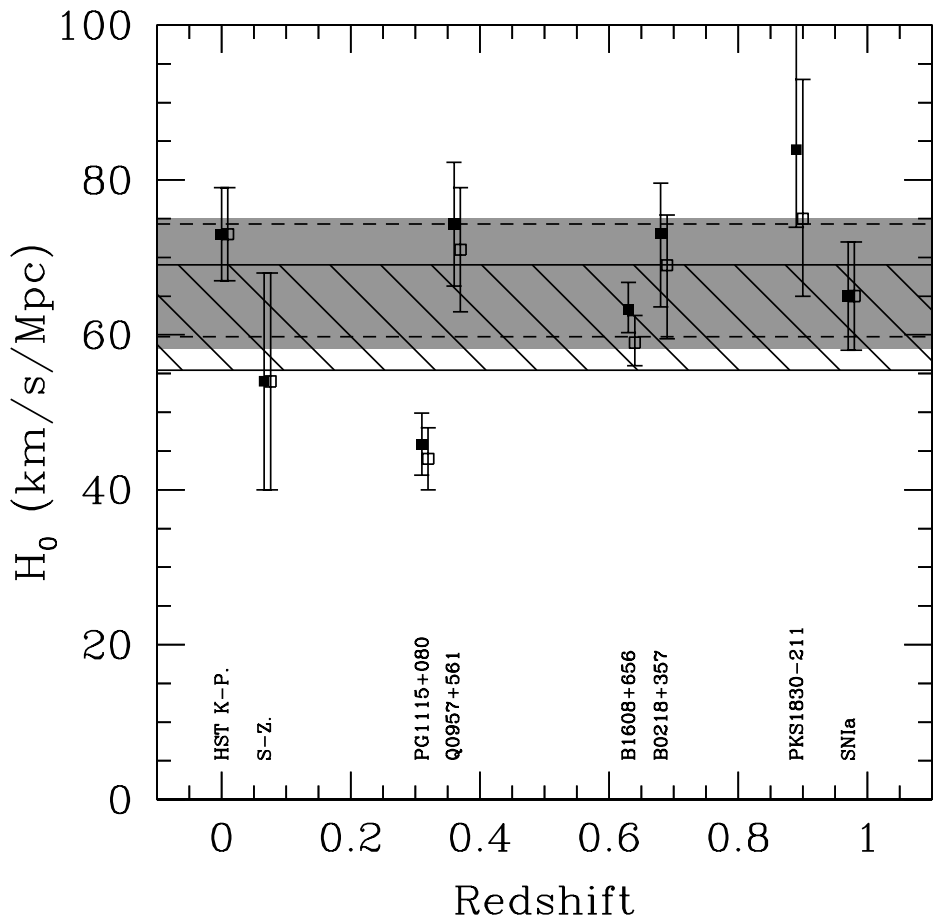}
}
\figcaption{\small The Hubble parameter determined on cosmological
scales plotted as function of redshift. The average local value,
presented by Freedman et al.\ (1998; HST Key Project for the
Extra-galactic Distance Scale) has a redshift $z=0$. The SNIa
determination has been placed on the highest redshift SNIa in the
sample (Riess et al.\ 1998). The redshift for the S-Z determination of
H$_0$ is the average cluster redshift in the sample (Myers et
al. 1997). The gravitational lenses have been placed at the primary
lens redshift.  The open symbols are the values of H$_0$ for
$\Omega_{\rm m}=1$ and $\Omega_{\Lambda}=0$, the closed symbols for
$\Omega_{\rm m}=0.3$ and $\Omega_{\Lambda}=0.7$. The local, S-Z and
SNIa determinations are almost independent of $\Omega_{\rm m}$ and
$\Omega_{\Lambda}$. The densily shaded region indicates the
$\pm$2--$\sigma$ region around the average of H$_0$
determined from the \sa~lens sample for $\Omega_{\rm m}=0.3$ and
$\Omega_{\Lambda}=0.7$. The dashed horizontal lines indicate the
$\pm$2--$\sigma$ region around the average of H$_0$
determinated from the \sa~lens sample plus the S-Z, SNIa and local
determinations. The lower shaded region indicates the $\pm$2--$\sigma$
region around the average of the \sa~lens sample for
$\Omega_{\rm m}=1.0$ and $\Omega_{\Lambda}=0.0$. The 1--$\sigma$
error-bars indicate the statistical errors.}
\end{center}
\end{figure}

The agreement on H$_0$ from four out of five gravitational lens
systems, the SNIa , the S-Z and local determinations is also an
independent confirmation that the lens mass models should not deviate
too much from isothermal, as was already shown from some of the
individual gravitational lens systems.  If we combine the strong
dependence on $\gamma$ of H$_0$ from B1608+656 (Fig.2) with the demand
that H$_0^{\rm GL}$ should be between the local and SNIa
determinations, then on average $\gamma$ for the \sa~sample should not
deviate by more than about 10 per cent from $\gamma$=0.50 (isothermal)
for $\Omega_{\rm m}=0.3$ and $\Omega_{\Lambda}=$0.0--0.7.

\section{Conclusions}

Using VLBA imaging, VLA monitoring and {\it Hubble Space Telescope}
WFPC2 and NICMOS imaging data, we constrained mass models of the lens
galaxies in the gravitational lens system B1608+656. The best mass
model gives (i) an agreement between the observed and model image
positions well within the observational errors, (ii) the radio flux
density ratios of the images to within 10\% (except for the faintest
radio image) and (iii) the observed time delays (Paper I) to within
1\%, although this small scatter is a coincidence.

Using the three time delays from B1608+656 (Paper I), the best
isothermal mass model gives
$$ 
     {\rm H}_0^{\small 1608}=59^{+7}_{-6}\mbox{ km s}^{-1}\mbox{ Mpc}^{-1}
     \;\;(\Omega_{\rm m}=1,\;\Omega_{\Lambda}=0),
$$ 
with 2--$\sigma$ errors. When $\Omega_{\rm m}=0.3$ and
$\Omega_{\Lambda}=$0.0--0.7, H$_0=(65-63)^{+7}_{-6}$ km
s$^{-1}$ Mpc$^{-1}$. All models give robust values for H$_0$, but a
systematic error of $\pm15$ km s$^{-1}$ Mpc$^{-1}$ remains, due
to a 20\% uncertainty in the radial mass profile.

Also under the explicit assumption of the isothermal mass model, we
determine the average of H$_0$ from a sample of four lenses
(B0218+357, Q0957+561, B1608+656 and PKS1830-211). We find
$$ 
    {\rm H}_0^{\rm GL}=69\pm7\mbox{ km s}^{-1}\mbox{ Mpc}^{-1}
    \;\;(\Omega_{\rm m}=1,\;\Omega_{\Lambda}=0),
$$ 
with 2--$\sigma$ errors. For $\Omega_{\rm m}=0.3$ and
$\Omega_{\Lambda}=$0.0--0.7 this increases to ${\rm H}_0^{\rm
GL}=74\pm8 \mbox{ km s}^{-1}\mbox{ Mpc}^{-1}$.  When including
PG1115+080, these values decrease to 64$\pm$11 km s$^{-1}$ Mpc$^{-1}$
and 68$\pm$13 km s$^{-1}$ Mpc$^{-1}$, respectively.

These values agree very well with the local (Freedmann et al.\ 1998),
Type Ia Supernovae (Riess et al.\ 1998) and S-Z (Myers et al.\ 1997)
determinations, supporting the reliability of the cosmological
determinations of H$_0$ from gravitational lenses. On average the lens
determinations agree best with local determinations for a low density
universe.

Moreover, we find that the mass model of B1608+656 is close to
isothermal ($\gamma=0.50\pm0.10$), in good agreement with Q0957+561
and MG1654+1346.  The close agreement on H$_0$ from four out five
gravitational lens systems and the agreement on $\gamma$ from three
gravitational lens systems suggest the existence of a `universal' mass
profile that can describe the mass distribution of spirals,
ellipticals and cluster ellipticals. This profile must on average be
very close to isothermal, perhaps following the profile found by
Navarro, Frenk \& White (1996), which is indeed close to isothermal in
the intermediate region probed by lensing.

\acknowledgments

We like to thank Phillip Helbig, Ger de Bruyn \& Steve Myers for
giving useful comments and suggestions to improve this paper. We also
like to thank Chris Lidman for providing the redshift of PKS1830-211
prior to publication.  LVEK acknowledges the support from an NWO
program subsidy (grant number 781-76-101). CDF acknowledges the
support from an NSF grant, \#AST 9420018. This research was supported
in part by the European Commission, TMR Programme, Research Network
Contract ERBFMRXCT96-0034 `CERES'.

\end{document}